\documentstyle[aps,pra,preprint,tighten]{revtex}

\def\ket#1{| #1 \rangle}

\def\expect#1{\langle #1 \rangle}
\def\ahat{{\hat a}}
\def\adag{{\hat a}^\dagger}
\def\chat{{\hat c}}
\def\cdag{{\hat c}^\dagger}
\def\E{{\hat E}}
\def\Ep{{{\hat E}^+}}
\def\Em{{{\hat E}^-}}
\def\re{{\rm Re}}
\def\im{{\rm Im}}
\def\e{{\rm e}}

\begin{document}

\title{Interference in dielectrics and pseudo-measurements}

\author{Todd A. Brun\thanks{Current address:  Institute for Theoretical
Physics, University of California, Santa Barbara, CA  93106-4030 USA} \\
Physics, Queen Mary and Westfield College \\
University of London, London  E1 4NS, United Kingdom \\
and \\
Stephen M. Barnett \\
Department of Physics and Applied Physics, \\
University of Strathclyde, \\
107 Rottenrow, Glasgow  G4 0NG, United Kingdom}

%\date{\today}
\date{December 6, 1996}

\maketitle

\begin{abstract}
Inserting a lossy dielectric into one arm of an interference experiment
acts in many ways like a measurement.  If two entangled photons are
passed through the interferometer, a certain amount of information is gained
about which path they took, and the interference pattern in a coincidence
count measurement is suppressed.
However, by inserting a {\it second} dielectric
into the other arm of the interferometer, one can restore the interference
pattern.  Two of these pseudo-measurements can thus cancel each other out.
This is somewhat analogous to the proposed quantum eraser experiments.
\end{abstract}

\section{Introduction}
In recent years, experimenters have gained an unprecedented ability to
perform experiments on single, microscopic quantum systems: individual atoms
or ions in traps, photons in optical systems, electron pairs in
Josephson junctions.  One benefit of this increased experimental power is
the ability to probe the physics of measurement itself.

Bohr formulated quantum mechanics solely in terms of the observation of
quantum systems by classical measurement devices.  To him, a quantum
measurement must be ``based on registrations obtained
by means of suitable amplification devices with irreversible functioning
such as, for example, permanent marks on a photographic plate, caused by
the penetration of electrons into the emulsion.  In this connection, it
is important to realize that the quantum-mechanical formalism permits
well-defined applications referring only to such closed phenomena''
\cite{Bohr}.  Thus, to measure a quantum system was to destroy it.

In the experiments of the day this was certainly true; but improvements
in experimental technique now make it possible to probe individual systems
without destroying them, or to make repeated or even continuous measurements
on a single microscopic system.

Jeffers and Barnett \cite{Jeffers} have described an
experimental set-up using a two-photon interferometer, based on earlier
proposals of Franson \cite{Franson} and Steinberg, Kwiat, and Chiao
\cite{Steinbergetal}.
Passing entangled pairs of photons through
this apparatus produces an interference pattern in the
measurement of coincidences.  If one inserts a lossy dielectric into
one arm of the interferometer, this interference pattern is substantially
destroyed, even though neither photon was absorbed by the dielectric.
In effect, {\it because} neither photon was absorbed, one has gained
information about which arm of the detector they passed through.
This set-up is described in section II.A.
(See figure 1.)

This situation is in many ways similar to an ordinary quantum measurement.
Information is gained about the state of the system; the superposition which
leads to the interference pattern is suppressed.  Moreover, this works by
permitting the photons to interact with an essentially classical, dissipative
environment with many internal degrees of freedom (the dielectric),
quite similar to a classical measuring device.  What is more, in this
interaction, the photon is not destroyed -- one keeps only the events in
which two photons were present (the coincidences).

Nevertheless, there are a number of important differences between this
and a measurement.  The superposition is only
modified, not destroyed; by an appropriate manipulation, coherence can
be retained, and the interference pattern restored.  We describe in section
II.B a related experiment, where by inserting
a second dielectric into the other arm of the interferometer we can
restore the interference pattern (though naturally the overall count
rate goes down).
Thus, a second of these ``pseudo-measurements'' can
undo the effects of the first.  (See figure 2.)

This experiment is similar to the proposed {\it quantum erasers}
\cite{Eraser}, in which interference is
destroyed by allowing the photons to interact with an auxiliary system
which stores ``which-way'' information.  In these erasers, the interference
pattern can be restored by performing an appropriate measurement on the
auxiliary system and plotting coincidences.

In section III we discuss the implications of this experiment for the
understanding of measurements, and its similarities to and differences from
the quantum eraser.

\section{The experimental set-up}

The entangled photon pair is produced by parametric down conversion
and passed through an interferometer (see figures 1 and 2).  If the
dielectrics are absent, one can adjust the lengths of the two arms of
the interferometer so that the electromagnetic pulses arrive at the
beam splitter simultaneously.  An overlap between the wave packets causes
them to tend to leave the beam splitter by the same port
\cite{Hongetal,Loudon,RarityTapster}.
If one then measures coincidences in the two detectors $D_a$ and $D_b$,
their rate falls off as the overlap of the wave packets improves.
It is possible to tune to a
dark fringe where the number of coincidences goes to zero.

The photon pair produced by parametric down-conversion may be 
described by the state
\begin{equation}
\ket\Psi = \int_0^\infty d\omega\ f_1(\omega) f_2(\omega)
  \adag_1(\omega) \adag_2(\Omega-\omega) \ket0,
\label{state}
\end{equation}
where $\adag_i(\omega)$ is the creation operator for a photon in arm $i$
with frequency $\omega$, and $\ket0$ is the vacuum state
of the free-space electromagnetic field.  The functions $f_1$ and $f_2$
are bandwidth functions, which we will assume to be narrow, and $\Omega$
is the frequency of the down converted photon.

If the two entangled photons pass into dielectrics instead, one replaces
the operators $\adag_i$ above with {\it polariton creation operators}
$\cdag_i(x,\omega)$ and the electromagnetic vacuum with
$\ket{0_{\rm diel}}$, the ground state of the polaritons within
the dielectric \cite{Jeffers,Huttner}.

If we pass the photons through a narrow-bandwidth filter we can specify
the functions $f_1$ and $f_2$ such that
\begin{equation}
f_1(\omega) f_2(\omega) = \exp\biggl[
  - {{(\omega - \Omega/2)^2}\over{B^2}} \biggr],
\label{bandwidth}
\end{equation}
where $B^2$ determines the frequency spread.  It then becomes a good
approximation to expand the wave vector $k(\omega)$ within the dielectric
about the frequency $\Omega/2$,
\begin{equation}
k(\omega) = k_{\Omega/2} + \alpha (\omega-\Omega/2)
  + \beta (\omega - \Omega/2)^2.
\label{dispersion}
\end{equation}
In a lossy medium, the constants $k_{\Omega/2}$, $\alpha$, and $\beta$
will be complex numbers; $\re(\alpha)$ is the inverse group velocity,
and $\re(\beta)$ is the dispersion, while the imaginary components determine
the linear and quadratic dependence of the loss.

\subsection{The single-dielectric experiment}

As shown in \cite{Jeffers}, the probability of detecting two
photons at $D_a$ and $D_b$ at times $t_a$ and $t_b = t_a + \tau$,
respectively, is
\begin{equation}
P(t_a,t_b) = \eta \expect{ \Em_a(t_a) \Em_b(t_b) \Ep_b(t_b) \Ep_a(t_a) },
\end{equation}
where $\eta$ is the (constant) detector efficiency, and $\Ep_a(t_a)$ is
the positive-frequency component of the electric field
at the detector $D_a$ at time $t_a$, and
similarly for the other electric field operators.  These fields will be
superpositions of the fields $\E_1$ and $\E_2$ from the first and
second arms of the interferometer, respectively.  A 50-50 beam splitter
will produce superpositions at $D_a$ and $D_b$ which are $\pi/2$ out of
phase with each other.

In terms of the operators $\E_1$ and $\E_2$ this joint probability is
\begin{eqnarray}
P(t_a,t_b) = && {\eta\over4} \expect{ ( \Em_1(t_a) - i \Em_2(t_a) )
  ( \Em_2(t_b) - i \Em_1(t_b) ) \nonumber\\
&& \times ( \Ep_2(t_b) + i \Ep_1(t_b) )
  ( \Ep_1(t_a) + i \Ep_2(t_a) ) }.
\end{eqnarray}
There is only one photon in each beam of 1 and 2, so we can drop the
terms proportional to $(\E_i)^2$, and are left with
\begin{eqnarray}
P(t_a,t_b) = && {\eta\over4} \expect{
  ( \Em_1(t_a) \Em_2(t_b) - \Em_2(t_a) \Em_1(t_b) ) \nonumber\\
&& \times ( \Ep_2(t_b) \Ep_1(t_a) - \Ep_1(t_b) \Ep_2(t_a) ) }.
\label{coincidence1}
\end{eqnarray}

The electric field operator leaving the dielectric in arm 1 is
given by \cite{Huttner}
\begin{equation}
\Ep_1(x_1,t) = C_1 \int_0^\infty d\omega_1 \chat_1(0,\omega_1)
  \e^{i(k_1(\omega_1) x_1 - \omega_1 t)}
\label{dielectric_field}
\end{equation}
plus a Langevin term which vanishes when applied to $\ket{0_{\rm diel}}$,
the ground state of the dielectric.  $C_1$ is a complex constant.
The field operator in arm 2 is
\begin{equation}
\Ep_2(x_2,t) = i C_2 \int_0^\infty d\omega_2
  \ahat_2(\omega_2) \e^{- i \omega_2 (t - x_2/c) },
\label{vacuum_field}
\end{equation}
where $C_2$ is a real constant.  The lengths of the two arms are $x_1$
and $x_2$, respectively.
Given the initial state (\ref{state})
(with $\adag_1$, of course, replaced by $\cdag_1$), we can calculate the
term
\begin{eqnarray}
&& \Ep_2(x_2,t_b)\Ep_1(x_1,t_a) = i C_1 C_2 \int_0^\infty d\omega
  f_1(\omega) f_2(\omega) \nonumber\\
&& \times \int_0^\infty d\omega_1 \int_0^\infty d\omega_2
  \e^{i(k_1(\omega_1) x_1 - \omega_1 t_a)}
  \e^{-i \omega_2(t_b - x_2/c)} \nonumber\\
&& \times \ahat_2(\omega_2) \chat_1(0,\omega_1)
  \cdag_1(0,\omega) \adag_2(\Omega-\omega)
  \ket{0_{\rm diel}}.
\end{eqnarray}
The usual delta-function commutators give
\begin{eqnarray}
&& \ahat_2(\omega_2) \chat_1(0,\omega_1)
  \cdag_1(0,\omega) \adag_2(\Omega-\omega)
  \ket{0_{\rm diel}} \nonumber\\
&& = \delta(\omega_1 - \omega) \delta(\omega_2 - \Omega + \omega)
  \ket{0_{\rm diel}}.
\end{eqnarray}
Similar expressions hold for the other terms in (\ref{coincidence1}).
Substituting the bandwidth function (\ref{bandwidth}) and the expression
for the wave vector (\ref{dispersion}), it is then possible to do the
integrals and evaluate the coincidence probability.

Let us assume that the detectors integrate over a time $T$, and let $T$
become very large.  Then if we integrate $t_a$ and $t_b$ over a full
range of times \cite{Jeffers}, we get a total coincidence probability of
\begin{equation}
P_c = \kappa\biggl( 1
  - \exp\biggl[ - {{(x_1 \im(\alpha_1))^2}
  \over{B^{-2} + 2 x_1 \im(\beta_1)}} \biggr]
  \exp\biggl[ - {{\tau_r^2}
  \over{B^{-2} + 2 x_1 \im(\beta_1)}} \biggr] \biggr),
\end{equation}
where $\tau_r$ is the time-delay between the two arms of the interferometer
\begin{equation}
\tau_r = x_2/c - \re(\alpha_1) x_1.
\end{equation}
$\re\alpha$ is the inverse group velocity 
associated with the dielectric in arm 1, as given in (\ref{dispersion}),
and $\kappa$ is an overall constant which reflects the probability of
no photon being absorbed by the dielectric.
By adjusting the lengths of the arms, it is possible to set $\tau_r = 0$.
In the absence of the dielectric (i.e., $\im(\alpha) = 0$) this is tuning to
a dark fringe; the coincidence probability becomes identically zero.  With
the dielectric present, however, there is always a nonzero probability of a
coincidence.
This reduction of the interference pattern by
interaction with a lossy medium is very similar to the effect of measuring
one path of a two-slit experiment, or to the destruction of interference
in the quantum eraser due to interaction with a system storing
``which-way'' information.

\subsection{The two-dielectric experiment}

Suppose now that we insert a second dielectric into the other arm of the
interferometer, as in figure 2.  How does this modify the probability of
coincidences?  The derivation is almost identical to that before.  In the
initial state (\ref{state}) both of the electromagnetic creation operators
$\adag_i(\omega_i)$ are replaced by polariton creation operators
$\cdag_i(0,\omega_i)$; the expression (\ref{vacuum_field}) for the electric
field in arm 2 is replaced by an expression analogous to
(\ref{dielectric_field}).  A dispersion relation (\ref{dispersion}) holds
for both arms.  Carrying out the integrals as before, we find that the
probability of coincidences becomes
\begin{eqnarray}
P_c = && \kappa' \biggl( 1 - \exp\biggl[
  - {{(x_1 \im(\alpha_1) - x_2 \im(\alpha_2))^2}
  \over{B^{-2} + x_1\im(\beta_1) + x_2\im(\beta_2)}} \biggr] \nonumber\\
&& \times \exp\biggl[ - {{\tau_r^2}
  \over{B^{-2} + x_1\im(\beta_1) + x_2\im(\beta_2)}} \biggr] \biggr),
\end{eqnarray}
where $\kappa'$ is now an overall constant representing the probability
of {\it neither} photon being absorbed, and $\tau_r$ becomes
\begin{equation}
\tau_r = x_2 \re(\alpha_2) - x_1 \re(\alpha_1).
\end{equation}
Here we see that the visibility now depends on the {\it difference} between
the absorption in the two dielectrics at the frequency $\Omega/2$.  By
choosing the arm lengths $x_1$ and $x_2$ and the linear absorptions
$\im(\alpha_1)$ and $\im(\alpha_2)$ appropriately, we can restore the dark
fringe, resurrecting the interference pattern.  Thus, adding a second
``pseudo-measurement'' undoes the effect of the first, just as inserting
an appropriate measurement of the {\it ancilla} in the quantum eraser
can restore the interference effect.

\section{Discussion}

Great care in interpretation is required as experiments begin to probe
individual quantum systems nondestructively.  It was possible to reconstruct
the interference pattern in section 2b, because although the interference
pattern had been exponentially suppressed, it had been suppressed
coherently.  By selectively suppressing the other component of the wavefunction
the pattern could be restored,
though at a large cost in the overall count rate.

An analogy to this might be the two slit experiment.  Suppose one could
narrow one of the two slits, so that fewer particles could pass through
it relative to the other.  One would have gained some information about
the system---particles are more likely to pass through slit 2 than slit
1---and the interference pattern on the screen would be greatly reduced.
It could be restored, however, by narrowing slit 2 as well, reinstating the
full contrast between bright and dark fringes, at the cost of a large reduction
in the brightness of the whole pattern.

This type of experiment examines one piece of the process of measurement,
but only one.  Measurement involves the correlation of a quantum system
with the many degrees of freedom of a classical object, such as a measuring
device.  In the case of this experiment, this role is played by the internal
degrees of freedom of a lossy dielectric \cite{Jeffers}.  But more than this
is required:  the information transfered to these internal degrees of freedom
must be unrecoverable.

Certainly this is so if one of the photons is actually absorbed.  It is in
principle possible to tell whether a photon was absorbed or not; one could
measure the energy absorbed by the dielectric, for example.  The detailed
information of the photon's phase and correlations, however, would be lost.

In this experiment, however, we are looking at the case where the photon
escapes unscathed.  Though the amplitude of the component passing through
the dielectric is greatly diminished, it is still present, and retains
all the relevant information about the quantum state.  Thus, suitable
manipulations can recover this information and demonstrate the survival
of its correlations.

This situation differs somewhat from the quantum eraser \cite{Eraser}.
In that experiment,
the photons interact with an auxiliary system with a low number of degrees
of freedom.  This interaction, in effect, measures which path the photons
take.  The entanglement with the auxiliary system destroys the
interference pattern, but there is no loss of coherence; by combining the
photon measurement with an appropriate measurement of the auxiliary system,
it is possible to fully restore the interference fringes.

The two experiments have points in common as well, of course.
In each case, an interaction is performed which captures
one aspect of the measurement process, but which is
reversible by a second interaction.  This reversibility
in both cases arises because the information which has been extracted
from the quantum system is retained coherently---in the case of the
dielectric in the diminished (but not destroyed) component of the wavefunction,
and in the case of the quantum eraser in the entanglement with the
{\it ancilla}.  A measurement, in Bohr's sense, must be irreversible
to be final.

\section*{Acknowledgments}
This work was done at the Ecole d'Et\'e de Physique Th\'eorique,
Les Houches, France.
TAB would like to thank Ian Percival and Nicolas Gisin for useful
conversations.  SMB would like to thank Howard Wiseman for discussions
on quantum erasers.  Financial support was provided by the
EPSRC in Britain.

\vfil

Figure 1.  The experimental set-up for the single-dielectric interferometer.

\vfil

Figure 2.  The experimental set-up for the two-dielectric interferometer.

\vfil

\end{document}